# Multicriteria Analysis of Decentralized Wastewater Treatment Technologies for the Philippines

Egberto F Selerio Jr [a*]

[a] School of Engineering, University of San Carlos, Talamban, Philippines
[*] Corresponding author. E-mail: 20102931@usc.edu.ph

## ABSTRACT

This research focuses on decentralized wastewater treatment (DEWAT) technologies for the Philippines – motivated by the limited suitable wastewater treatment infrastructure in the country. A multi-criteria analysis (MCA), using the Analytic Hierarchy Process (AHP) and Delphi method, was employed to evaluate DEWAT technologies based on life cycle costs and wastewater treatment efficiency parameters such as $COD_t$, $BOD_5$, TSS, $NH_4$-N, TP, and hydraulic retention time. A two-factor Analysis of Variance (ANOVA) without replication was used to assess statistical differences between technologies. The analysis revealed that the Downflow Hanging Sponge (DHS) filter, Multi-Soil Layering (MSL) systems, and Moving Bed Biofilm Reactors (MBBRs) are the top-performing technologies, with no statistically significant differences in their overall performance. The DHS filter ranked highest, excelling in energy efficiency and nutrient removal, making it ideal for resource-scarce environments. MSL systems were noted for their broad-spectrum contaminant removal, while MBBRs demonstrated flexibility and scalability for semi-urban areas. A thorough analysis is carried out for these DEWAT technologies and insights for applicability in the Philippine context are provided.

## HIGHLIGHTS

- The study uses MCA with AHP to systematically evaluate DEWAT technologies for the Philippines – the first in the literature.
- Expert consensus through the Delphi method refines judgments on key criteria for technology selection.
- A two-factor ANOVA evaluates statistical differences between wastewater treatment technologies.

- DHS filters, MSL systems, and MBBRs are found most suited for specific environmental and geographical needs in the Philippines.
- Policy insights in the Philippine context are provided for the application of these technologies.

## 1  INTRODUCTION

Water and wastewater management remains a critical challenge across the globe, especially in densely populated regions and countries facing rapid urbanization. As cities grow, the demand for efficient wastewater treatment systems becomes more pressing. In many regions, centralized wastewater treatment has proven effective in managing population-dense urban areas, but these systems are often met with financial, infrastructural, and logistical barriers in resource-limited or geographically dispersed countries. The challenges are compounded by climate change, which affects water availability and heightens the need for resilient, sustainable solutions in countries vulnerable to extreme weather events. This problem is especially true to developing archipelagic countries around the world, but the specific case of the Philippines is taken in this research as a case study.

The Philippines is an archipelago with 7,641 islands, making it the 7th country with the most islands in the world, although it only has about 298,170 km$^2$ of land (Lang'at Jr., 2018). In the Philippine context, centralized domestic wastewater treatment is only practical in cities where centralized sewer lines can be built near the serviced population. However, cities in the Philippines are already very dense, which heavily hampers centralized treatment. The world's top three densest cities (by population) are in the Philippines, i.e., Manila, Pateros, and Mandaluyong (*Population Density by City 2023*, n.d.). In these cities, installing infrastructure to collect domestic wastewater for centralized treatment is extremely challenging for two main reasons. Firstly, there is already very little available space in these dense cities. Second, most of these cities have grown unplanned, so paths for building sewer lines are seldom available. In the rural island regions of the country, centralized treatment is impractical because hundreds, if not thousands, of such facilities, need to be built to cater to all the islands in the country – some of which are extremely remote. On top of infrastructural challenges, the cost of such a project will be extremely high. In the Philippines, which is a developing nation, the funds for such a project are unavailable because of more urgent issues like agriculture, healthcare, education, and housing, which take priority.

Decentralized wastewater treatment (DEWAT) technologies were recently viewed by several scholars as a suitable option to address the Philippine context (Damalerio et al., 2022; Monjardin et al., 2021; Velasco et al., 2023). In fact, several DEWAT technologies have been applied in the country. Some of them are nature-based technologies like constructed wetlands (Velasco et al., 2023), small-scale rotating biological contactors (Cabije et al., 2009), and anaerobic reactors treating slaughterhouse wastewater (Parker, 2011). DEWAT is advantageous to the Philippine context for several reasons. The vast number of islands and dispersed populations make centralized wastewater treatment impractical, while DEWAT serves small communities effectively (Parker, 2011). Rapid urban growth and high population density often outpace infrastructure development, but DEWAT can be incrementally implemented (Damalerio et al., 2022). Economic constraints and high costs of centralized systems make the more affordable DEWAT systems appealing (Damalerio et al., 2022). Additionally, DEWAT's modular scalability suits varying community sizes, and local community involvement in DEWAT fosters better maintenance and sustainability (Nasr et al., 2022). Lastly, DEWAT systems offer a variety of technological options that can be tailored to specific local conditions and resource availability (Bright-Davies et al., 2015).

Although some DEWAT technologies have been recognized and studied in the Philippines, their adoption remains limited (Damalerio et al., 2022). A densely populated archipelagic nation like the Philippines is highly exposed to the effects of climate change (Selerio et al., 2020). The recent El Niño phenomenon that occurred from March to May 2024 has exacerbated water scarcity by drying up dams in various regions, especially the Visayas, highlighting the urgent need for resilient water management solutions. As an archipelago, the Philippines faces unique vulnerabilities to climate-induced water crises due to its dispersed geography and reliance on local water resources (Damalerio et al., 2022; Greenpeace, 2007; Selerio et al., 2020; Tabios III, 2020; Velasco et al., 2023). This makes the country particularly susceptible to fluctuations in water availability and the impacts of extreme weather events (Tabios III, 2020). Given these challenges, selecting appropriate wastewater treatment technology is crucial for the Philippines as a resource-limited and climate-vulnerable setting.

Furthermore, a comprehensive examination of DEWAT technologies is essential to direct engineering efforts and policies toward the most cost-effective and resilient solutions. Decentralized systems can offer flexible, localized treatment options that mitigate the environmental impact and provide sustainable water management in isolated and densely populated urban areas, which are abundant in the Philippines. To address the critical need for adaptive strategies, this study executes a multi-criteria analysis of applicable DEWAT technologies in the Philippine context. It considers the unique challenges and opportunities posed by the Philippines' status as a densely populated and climate-vulnerable island nation. While this analysis is case-specific, its results is significant to countries with similar situations such as Indonesia, Vietnam, and Thailand. These countries, like the Philippines, face challenges of rapid urbanization, climate change, and the need for sustainable water management in both urban and rural areas. The shared geographical features of being archipelagic nations, coupled with their exposure to natural disasters like typhoons and floods, make decentralized wastewater treatment systems highly relevant. Additionally, these nations also experience varying levels of infrastructure development, which makes decentralized solutions more adaptable to their diverse urban and rural settings.

In this study, seven DEWAT technologies are systematically assessed. The Analytic Hierarchy Process (AHP) is used for the assessment together with the Delphi method. Ten quantitative and qualitative criteria are used which are divided into two sets: wastewater treatment criteria and technology life-cycle criteria. Statistical analysis is deployed on the AHP results and managerial insights from the results in the Philippine context are provided. The analysis employed here is the first systematic approach to analyze DEWAT technologies in the Philippine context, which is the main contribution of this work.

## 2 DEWAT TECHNOLOGIES

Conventional wastewater treatment technologies for DEWAT application include constructed wetlands, septic tanks, multi-soil-layering, sand filters, rotating biological contactors, moving bed biofilm reactors (Ulrich et al., 2009), and the DHS filter (Nasr et al., 2022). DEWAT designs of some anaerobic treatment technologies, i.e., anaerobic lagoons, up-flow anaerobic sludge blanket (UASB), anaerobic baffled reactors, and anaerobic membrane bioreactors (AMB), have high removal of organic matter and can produce biogas for energy or fuel source. However, these technologies are not considered in this study for three reasons. Firstly, all anaerobic treatment technologies exhibit low nutrient removal, which is important given the strict requirements of the Philippines's Department of Environment and Natural Resources (DENR Administrative Order No. 2016-08, 2016) for Class C effluent. Secondly, anaerobic microorganisms are inhibited by surfactants and other xenobiotic compounds from households (Khalil & Liu, 2021), which is prevalent, especially when treating greywater. Greywater treatment is among the most common DEWAT applications (Ghaitidak & Yadav, 2013). Third and lastly, they can require a large land area (i.e., anaerobic lagoons) or hefty technical skill (i.e., AMB) to operate, which makes them unattractive for DEWAT application in developing island nations like Philippines, which has limited land area, among other things. The only anaerobic treatment technology considered in this review is the septic tank, the most common yet controversial DEWAT technology globally (Brandes, 1978; Whelan & Titamnis, 1982).

## 2.1 CONSTRUCTED WETLAND

Constructed wetlands are engineered systems that mimic the natural processes occurring in wetlands. Wastewater flows through a bed of gravel or other porous material, where plants and microorganisms break down and remove contaminants. A summary of effluent characteristics treated by constructed wetland technologies reported in the review of Valipour & Ahn (2016) is presented in **Table 2-1**. Constructed wetlands effectively remove organic matter, nutrients, and pathogens from domestic wastewater. They also require minimal energy input. However, they typically require a large land area. Furthermore, their performance is affected by weather conditions and seasonal fluctuations.

**Table 2-1** Summary of effluent characteristics treated using conventionally constructed wetlands (Valipour & Ahn, 2016)

| Type[a,b] | Temp | pH | $COD_t$ | $BOD_5$ | TSS | $NH_4$-N | TP | HRT | Q | HLR |
|---|---|---|---|---|---|---|---|---|---|---|
| | °C | | mg/L | mg/L | mg/L | mg/L | mg/L | day | m$^3$/day | m$^3$/m$^2$/day |
| SF | 24-27 | 7.4-7.7 | 87.1-195 | 75.95-96.1 | 48 | 9.3 | - | 1.5-1.8 | 0.03 | 0.27 |
| HSF | 28–30 | 6.5-7.69 | 89-148 | 8.4-23.5 | 2.3-47 | 1.992-28.8 | 0.5-25.5 | 1-6 | 0.0125-2 | 0.014-0.34 |
| Min-Max[c] | 24-30 | 6.5-7.69 | 87.1-195 | 8.4-96.1 | 2.3-48 | 1.992-28.8 | 0.5-25.5 | 1.5-1.8 | 0.0125-2 | 0.014-0.34 |

[a]SF – Surface Flow, HSF – Horizontal Subsurface Flow, Min-Max – the minimum and maximum values; [b]Temp – Temperature, $COD_t$ – total COD, $BOD_5$ – 5-day BOD, TSS – total suspended solids, $NH_4$-N – ammonium + ammonia N species, TP – total P, HRT – hydraulic retention time, Q – volumetric flow rate, HLR – hydraulic loading rates; [c]Min-Max refers to the value range for each parameter in the table, which is obtained based on the minimum and maximum values presented in every column

## 2.2 SEPTIC TANK

Septic tanks are underground chambers where solids and liquids in wastewater are separated. Solids settle at the bottom and are partially decomposed by anaerobic bacteria. On the other hand, the liquid portion flows out to a drain field or leach field for further treatment. A summary of effluent characteristics treated by conventional septic tanks is presented in **Table 2-2**. Septic tanks are simple to construct and maintain, but they require periodic pumping to remove accumulated solids and can become a source of groundwater contamination if not properly managed.

**Table 2-2** Summary of effluent characteristics treated using conventional septic tanks

| Ref. | Temp | pH | $COD_t$ | $BOD_5$ | TSS | $NH_4$-N | TP | HRT | Q | HLR |
|---|---|---|---|---|---|---|---|---|---|---|
|  | °C |  | mg/L | mg/L | mg/L | mg/L | mg/L | day | m³/day | m³/m²/day |
| Whelan & Titamnis (1982) | 20 | 7 | - | 52-316 | 22-47 | 63-201 | 12.3-26 | 6 | 0.657 | - |
| Beavers and Gardner as cited in Beal et al. (2005) | - | - | - | 150-180 | 100-180 | - | 10-15 | - | - | - |
| Gardner et al. (2013) | 12-23[b] | 6.8-8[b] | - | 120-150 | 40-190 | 40-50 | 10-15 | - | - | - |
| Min-Max | 12-23 | 6-8 | - | 52-180 | 22-190 | 40-201 | 12.3-26 | 6 | 0.657 | - |

[a]Ref. – Reference; [b]From Ziebell et al. (1975) which is the data source of Gardner et al. (2013)

## 2.3 MULTI-SOIL LAYERING

Multi-soil-layering systems distribute wastewater through perforated pipes in trenches filled with gravel or other porous materials. The wastewater percolates through the soil, which acts as a natural filter, removing contaminants through physical, chemical, and biological processes. A summary of effluent characteristics treated by multi-soil-layering is presented in **Table 2-3**. Multi-soil-layering can effectively treat domestic wastewater but depend on suitable soil conditions and sufficient land area.

**Table 2-3** Summary of effluent characteristics treated using multi-soil-layering systems

| Ref. | Temp | pH | $COD_t$ | $BOD_5$ | TSS | $NH_4$-N | TP | HRT | Q | HLR |
|---|---|---|---|---|---|---|---|---|---|---|
| | °C | | mg/L | mg/L | mg/L | mg/L | mg/L | day | m$^3$/day | m$^3$/m$^2$/day |
| Zhang et al. (2015) | 9.8-23.5 | 6.62-7.02 | 17.92-31.18 | - | - | 2.91-4.82 | 0.31-0.38 | 1.8 | 0.0512 | 0.66 |
| Latrach et al. (2014) | 22-26 | 7.08-7.12 | 120-134 | 27-31 | 11.8-12.2 | 3.5-3.9 | 0.5-0.9 | - | - | 0.2 |
| Ho & Wang (2015) | - | 6.69-7.57 | 36.5-113.3 | - | 0.88-3.97 | 0.07-1.72 | 0.08-0.45 | 0.5 | 0.1 | 0.5-3 |
| Min-Max | 9.8-26 | 6.62-7.57 | 17.92-134 | 27-31 | 0.88-12.2 | 0.07-4.82 | 0.08-0.9 | 0.5-1.8 | 0.0512-0.1 | 0.5-3 |

## 2.4 SAND FILTER

Sand filter systems consist of a bed of sand or other granular material through which wastewater percolates. Contaminants are removed by physical filtration, adsorption, and microbial activity. Sand filters can achieve high levels of pollutant removal, but they may require periodic maintenance to prevent clogging and maintain performance. A summary of effluent characteristics treated by sand filter systems is presented in **Table 2-4**.

**Table 2-4** Summary of effluent characteristics treated using sand filter systems

| Ref. | Temp | pH | $COD_t$ | $BOD_5$ | TSS | $NH_4$-N | TP | HRT | Q | HLR |
|---|---|---|---|---|---|---|---|---|---|---|
| | °C | | mg/L | mg/L | mg/L | mg/L | mg/L | day | m³/day | m³/m²/day |
| Tyagi et al. (2009) | - | 7.1-7.8 | 15-27 | 3.6-11.5 | 11-30 | - | - | 0.35-0.71 | 0.02-0.05 | 3.36-6.24 |
| Katukiza et al. (2014) | 24.3-25 | 7.2-7.9 | 858.3-886.91 | 371.25-438.75 | 350-830 | 7.66-9.63 | 1.39-1.48 | - | - | 0.2-0.4 |
| Kaetzl et al., (2020) | 21-23 | 7.6 | 69-116 | - | - | 39-40 | 4.3-4.9 | 0.5-1.27 | 2.83E-4-7.2E-4 | 1.2 |
| Min-Max | 21-25 | 7.1-7.9 | 15-886.91 | 3.6-438.75 | 11-830 | 7.66-40 | 1.39-4.9 | 0.35-1.27 | 2.83E-4-0.05 | 0.2-6.24 |

## 2.5 ROTATING BIOLOGICAL CONTACTOR

Rotating biological contactors (RBCs) are a fixed-film biological treatment process that uses rotating discs to support biofilm growth. Wastewater flows through the RBC unit, and the rotating discs provide aeration and contact with the biofilm. RBCs can be sized for small-scale or DEWAT applications and are often used for individual households or small communities. A summary of effluent characteristics treated by conventional RBCs is presented in **Table 2-5**. A disadvantage of the RBC is its vulnerability to mechanical failure. Furthermore, the overgrowth of biofilm on the discs can lead to uneven biomass distribution, which requires periodic cleaning or removal of excess biomass.

**Table 2-5** Summary of effluent characteristics treated using conventional RBCs

| Ref. | Temp | pH | COD$_t$ | BOD$_5$ | TSS | NH$_4$-N | TP | HRT | Q | HLR |
|---|---|---|---|---|---|---|---|---|---|---|
| | °C | | mg/L | mg/L | mg/L | mg/L | mg/L | day | m$^3$/day | m$^3$/m$^2$/day |
| Waqas et al. (2020) | 30 | 6.69-7.02 | 56.7-105.3 | - | - | 0.06-0.08 | 0.17-0.21 | 0.36 | 0.018 | 0.081 |
| Waqas et al. (2021) | 30 | 6.65-6.93 | 53-97 | - | - | 0.04-0.06 | 0.15-0.19 | 0.36 | 0.018 | 0.081 |
| Waqas et al. (2022) | 30 | 6.79-6.85 | 70.7-85.7 | - | - | 0.02-0.04 | - | 0.38-0.63 | 0.067-0.112 | 2.973-4.956 |
| Maheepala et al. (2022) | 11.8-24.4 | 7.1-7.3 | 45-66 | 6-15 | - | 15.4-27.3 | - | 0.083 | 0.482 | - |
| Min-Max | 11.8-30 | 6.65-7.3 | 45-105.3 | 6-15 | - | 0.02-27.3 | 0.15-0.21 | 0.36-0.63 | 0.018-0.482 | 0.081-4.956 |

## 2.6 MOVING BED BIOFILM REACTOR

Moving bed biofilm reactors (MBBRs) are a type of biological wastewater treatment process that combines the advantages of suspended growth and attached growth systems. In MBBRs, specialized plastic biofilm carriers with a high surface area provide a surface for biofilm growth. These carriers are suspended and continuously mixed in an aerated reactor, allowing the biofilm to contact the wastewater. A summary of effluent characteristics treated by multi-soil-layering is presented in **Table 2-6**. A disadvantage of MBBRs is the aeration requirement, which is energy intensive.

**Table 2-6** Summary of effluent characteristics treated using MBBRs

| Ref. | Temp | pH | COD$_t$ | BOD$_5$ | TSS | NH$_4$-N | TP | HRT | Q | HLR |
|---|---|---|---|---|---|---|---|---|---|---|
| | °C | | mg/L | mg/L | mg/L | mg/L | mg/L | day | m$^3$/day | m$^3$/m$^2$/day |
| Li et al. (2022) | 14-22 | 7.21-8.43 | 23.64-39.48 | - | - | 2.11-2.84 | 0.23-0.53 | 0.33 | 36 | 6.82 |
| Typical values for full-scale WWTPs[a] | 20-30 | 6.9-7.4 | 50-75 | 25-32 | 15-22 | 7-10 | 1.5-2.1 | 0.7-1.2 | 100-130 | 10-14 |
| Min-Max | 14-30 | 6.9-8.43 | 23.64-75 | 25-32 | 15-22 | 2.11-10 | 0.23-2.1 | 0.33-1.2 | 36-130 | 6.82-14 |

[a]Summary from the following review papers: di Biase et al. (2019), Leyva-Díaz et al. (2020), Saidulu et al. (2021). Data are chosen from studies reporting Q<150 m$^3$/day, which is the criteria used for classifying a small-scale WWTP

## 2.7 DOWN-FLOW HANGING SPONGE FILTERS

The DHS filter is an aerobic attached-growth process in which wastewater flows downward through a column filled with sponge-like materials that provide a surface for bacterial growth, which compose the biofilm. It is a trickling filter that uses sponge-like materials as biofilm carriers. DHS systems have shown promise as a low-cost and energy-efficient technology for DEWAT (Maharjan et al., 2020). They can achieve high removal rates of organic matter, suspended solids, and nutrients and require relatively small land areas compared to other technologies. A summary of effluent characteristics treated by the pilot-scale DHS filters, as reviewed by Tyagi et al. (2021), is presented in **Table 2-7**. The performance of DHS filters is reduced by uneven wastewater flow through the biofilm carriers (Tyagi et al., 2021).

**Table 2-7** Summary of effluent characteristics treated using pilot-scale DHS filters

| Ref. | Temp | pH | COD$_t$ | BOD$_5$ | TSS | NH$_4$-N | TP | HRT | Q | HLR |
|---|---|---|---|---|---|---|---|---|---|---|
| | °C | | mg/L | mg/L | mg/L | mg/L | mg/L | Day | m$^3$/day | m$^3$/m$^2$/day |
| Min-Max | 9-33[b] | 6.5-7.6[d] | 32-69[a] | 2-30[a] | 8-32[a] | 28-100[a] | - | 0.06-0.35[c] | 0.29-4.6[b] | - |

[a]From table 2 of Tyagi et al. (2021); [b]From table 4 of Tyagi et al. (2021); [c]From the discussions of Tyagi et al. (2021); [d]From various cited papers reviewed by Tyagi et al. (2021)

## 3 MATERIALS AND METHOD

Investment in DEWAT requires developing and implementing appropriate technologies, supportive policies, and capacity building to ensure successful adoption and operation. Firstly, an appropriate technology must be chosen. A multi-criteria analysis (MCA) using the Analytic Hierarchy Process (AHP) is used to determine the most appropriate technology in the Philippine context. For brevity, the mathematical details of AHP are omitted here and the reader is instead referred to Dos Santos et al. (2019). The MCA considers the following criteria: capital investment, capital replacement, electricity, operation and maintenance; effluent COD$_t$, BOD$_5$, TSS, NH$_4$-N, and TP; and HRT. The consideration of the first four criteria is based on the life cycle cost (LCC) assessment of (Rawal & Duggal, 2016) with the omission of residual value and revenue because it is assumed that the assessed technologies have zero value at the end of their operational life and are not revenue generating. This assumption is held to take the context of the Philippine government spearheading the preliminary DEWAT implementations in the country, which means that such operations is paid for with tax money and is not for profit. The rest of the criteria are considered because they are the most common wastewater treatment performance parameters.

### 3.1 THE ANALYTIC HIERARCHY PROCESS (AHP)

The AHP method is a structured technique for organizing and analyzing complex decisions, involving multiple criteria. The following steps were used in the AHP analysis:

1. Define the Goal, Criteria, and Alternatives: The goal is to select the most appropriate DEWAT technology for the Philippine context. The criteria included capital investment, capital replacement, electricity, operation and maintenance, effluent $COD_t$, $BOD_5$, TSS, $NH_4$-N, TP, and HRT. These criteria were based on life cycle cost (LCC) assessments (Rawal & Duggal, 2016) and the need to meet the Philippines' Class C effluent standards (DENR Administrative Order No. 2016-08, 2016). The seven DEWAT technologies presented in the previous section were evaluated.
2. Pairwise Comparisons: Each criterion was compared against every other criterion by the experts. The Delphi method is used to pool the experts' judgments; the details of which are discussed in the next section. For the comparison, the data on effluent concentrations of $COD_t$, $BOD_5$, TSS, $NH_4$-N, and TP that is presented in the previous section is opted for instead of their corresponding removal percentages because the technology selection is conducted on the premise of determining the most appropriate technology for achieving the Philippine's Class C effluent standards (DENR Administrative Order No. 2016-08, 2016). Since the standards are stated in terms of the compounds' effluent concentration, i.e., mg/L, it is appropriate that the MCA of the alternative technologies be conducted in the same units. The mid-range values of $COD_t$, $BOD_5$, TSS, $NH_4$-N, TP, and HRT are used for the assessment, while capital investment, capital replacement, electricity, operation, and maintenance are qualitatively assessed for each DEWAT technology due to a lack of data in the current literature. The qualitative assessment is performed using a five-point scale, with the scoring description that was provided to the experts presented in **Table 3-1**. A matrix was developed where each cell represented the relative importance of criterion over criterion. For this study, criteria weights were held equal, as agreed by the expert decision makers – implying an equal importance assumption.

**Table 3-1**. Scoring scale and description

| Scoring scale | Description |
|---|---|
| 1 | [Life cycle criteria] of [DEWAT technology $i$] is equal to that of [DEWAT technology $j$], but they are almost the same |
| 2 | [Life cycle criteria] of [DEWAT technology $i$] is more than that of [DEWAT technology $j$], but they are almost the same |
| 3 | [Life cycle criteria] of [DEWAT technology $i$] is slightly more than that of [DEWAT technology $j$] |
| 4 | [Life cycle criteria] of [DEWAT technology $i$] is moderately more than that of [DEWAT technology $j$] |
| 5 | [Life cycle criteria] of [DEWAT technology $i$] is exceedingly more than that of [DEWAT technology $j$] |

3. Calculation of Priority Weights: The pairwise comparison matrix was normalized, and the priority weights for each criterion were calculated by averaging across each row of the normalized matrix.
4. Consistency Check: A consistency ratio (CR) was calculated to ensure that the judgments were consistent. A CR value below 0.1 was considered acceptable, indicating the comparisons were consistent.

### 3.2 THE DELPHI METHOD

The assessment in the pairwise comparison step of the AHP method was performed using the Delphi method following Skulmoski et al. (2007). The method was used to qualitatively assess the life-cycle criteria for which there are limited available quantitative data (i.e., capital investment, capital replacement, electricity, operation and maintenance). The following steps were involved:

1. Selection of Experts: Four water management experts, as identified in Selerio et al. (2020), were chosen for this iterative assessment. The selected experts possessed significant experience in wastewater management and were familiar with decentralized wastewater treatment technologies, which ensured that their insights were highly relevant and informed. The meeting with the experts was conducted online.
2. Round One - Initial Ratings: The scoring description in **Table 3-1** is used and each expert was asked to provide initial pairwise ratings of the criteria for pairs of DEWAT technologies. This stage allowed for the identification of initial positions and perspectives.
3. Round Two - Feedback and Revision: After the initial ratings were gathered, the aggregated results were shared anonymously with all participating experts. Each expert was given the opportunity to revise their ratings based on the summary of the group's responses. This iterative process encouraged experts to reconsider their assessments and converge towards a common understanding. Experts were also encouraged to provide justifications for their ratings, which helped clarify differing viewpoints.
4. Subsequent Rounds and Iteration: The Delphi process continued through multiple rounds until a consensus was reached. Each subsequent round consisted of providing feedback summaries and allowing the experts to adjust their responses further. This step ensured that outlier opinions were brought in line with the general consensus, and any disagreements were minimized.

After multiple rounds, a final consensus was achieved on each criterion for each pair of DEWAT technologies and the results are recorded.
5. Data Analysis and Application: The consensus ratings obtained from the Delphi method were incorporated into the MCA framework as the pairwise comparison values for the life-cycle criteria and the rest of the AHP calculations proceed after. The datasets, the calculation, the files used for the calculation are published separately in Mendeley Data[*].

## 3.3 ANALYSIS OF VARIANCE (ANOVA)

Lastly, using a two-factor Analysis of Variance (ANOVA) without replication at a 0.05 level of significance, with the technologies as the factor and the criteria as the levels, the statistical significance of the results across the technology types is assessed. The ANOVA steps included:

1. Setting Up the ANOVA Table: The DEWAT technologies were treated as the factor. The criteria were treated as the levels.
2. Data Collection: The mid-range values of each water treatment criterion ($COD_t$, $BOD_5$, TSS, $NH_4$-N, TP, and HRT) and the consensus ratings from the Delphi method were used for the analysis.
3. Calculations: Calculation of the sum of squares, mean squares and F-value are carried out.
4. Decision Making: The calculated F-value was compared to the critical F-value at a 0.05 significance level. If the calculated F-value was greater, the null hypothesis (that there was no significant difference between the technologies) was rejected, indicating significant differences between technologies.

The ANOVA process is repeated on the top three DEWAT technologies based on the aggregate rank obtained from the AHP analysis.

## 4 RESULTS AND DISCUSSION

The results of the MCA are summarized in **Table 4-1**. Based on the criteria, the analysis elucidated the DHS filter as the top-most ranked DEWAT technology among the considered alternatives.

---

[*] The step-by-step procedures of the MCA, its raw data, as well as the calculation is available in a supplementary file accessible through the following link: https://doi.org/10.17632/wgycpw7byv.1

**Table 4-1** Summary of the total normalized score (TNS) of each alternative DEWAT technology

| Technologies | Criteria Number[a] | | | | | | | | | | TNS | Rank |
|---|---|---|---|---|---|---|---|---|---|---|---|---|
| | 1 | 2 | 3 | 4 | 5 | 6 | 7 | 8 | 9 | 10 | | |
| Constructed wetlands | 0.29 | 0.29 | 0.07 | 0.08 | 0.14 | 0.11 | 0.04 | 0.06 | 0.30 | 0.15 | 0.1624 | 5 |
| Septic tanks | 0.28 | 0.29 | 0.07 | 0.08 | 0.15 | 0.25 | 0.15 | 0.49 | 0.44 | 0.54 | 0.2424 | 7 |
| Multi-soil-layering | 0.14 | 0.14 | 0.13 | 0.08 | 0.08 | 0.06 | 0.01 | 0.01 | 0.01 | 0.10 | 0.0930 | 2 |
| Sand filter | 0.06 | 0.06 | 0.13 | 0.23 | 0.47 | 0.47 | 0.60 | 0.10 | 0.07 | 0.07 | 0.1907 | 6 |
| RBCs | 0.09 | 0.07 | 0.27 | 0.23 | 0.08 | 0.02 | 0.14 | 0.06 | 0.00 | 0.04 | 0.1229 | 4 |
| MBBRs | 0.06 | 0.07 | 0.20 | 0.23 | 0.05 | 0.06 | 0.03 | 0.02 | 0.03 | 0.07 | 0.1013 | 3 |
| DHS filter | 0.07 | 0.07 | 0.13 | 0.08 | 0.04 | 0.03 | 0.03 | 0.26 | 0.14 | 0.02 | 0.0873 | 1 |

[a]1-capital investment, 2-capital replacement, 3-electricity, 4-operation and maintenance, 5-$COD_t$, 6-$BOD_5$, 7-TSS, 8-$NH_4$-N, 9-TP, and 10-HRT

## 4.1 DISCUSSION BY DEWAT TECHNOLOGY

### CONSTRUCTED WETLANDS (TNS: 0.1624, RANK: 5)

Constructed wetlands, with a TNS of 0.1624 and ranked fifth, show a balance of moderate performance across various criteria. High capital investment and replacement costs (0.29 for both) reflect the substantial infrastructure needed for setup and periodic renewal. However, low electricity (0.07) and operation and maintenance (0.08) scores suggest energy efficiency and ease of upkeep, aligning with the minimal operational demands typically observed in passive treatment systems like surface flow (SF) and horizontal subsurface flow (HSF). Effluent characteristics show $COD_t$ levels (87.1-195 mg/L) and $BOD_5$ (8.4-96.1 mg/L) within acceptable ranges, explaining the average scores for these parameters (0.14 and 0.11, respectively). Constructed wetlands excel in TSS removal (score 0.04), which corresponds to lower TSS values in the effluent (2.3-48 mg/L), supporting their effectiveness in solids reduction. $NH_4$-N and TP scores (0.06 and 0.30, respectively) suggest variability in nutrient removal, which mirrors the broad range of effluent concentrations for NH4-N (1.992-28.8 mg/L) and TP (0.5-25.5 mg/L). Lastly, a moderate HRT score (0.15) reflects a controlled hydraulic retention time, which is crucial for effective treatment, as seen in the actual HRT values (1-6 days). These insights collectively explain the wetland's mid-range ranking.

### SEPTIC TANKS (TNS: 0.2424, RANK: 7)

Septic tanks, with a TNS of 0.2424 and ranked seventh, exhibit relatively high scores across most criteria, reflecting their limitations as a DEWAT technology. Capital investment (0.28) and replacement (0.29) scores align with their simple infrastructure, though long-term upkeep can be costly. Low electricity (0.07) and operation and maintenance (0.08) scores indicate minimal energy and labor demands, typical for passive systems. However, effluent characteristics like high $BOD_5$ (52-316 mg/L) and $NH_4$-N (63-201 mg/L) explain higher scores for $BOD_5$ (0.25) and $NH_4$-N (0.49), highlighting poor nutrient and organic matter removal. High TP concentrations (12.3-26 mg/L) correlate with a high TP score (0.44), indicating ineffective phosphorus reduction. The extended HRT of 6 days is reflected in the high HRT score (0.54), suggesting that even with prolonged retention times, septic tanks struggle to produce consistently low pollutant levels.

### MULTI-SOIL-LAYERING (TNS: 0.0930, RANK: 2)

Multi-soil-layering systems, with a TNS of 0.0930 and ranked second, demonstrate high overall efficiency. Low capital investment (0.14) and replacement (0.14) scores align with the relatively simple materials required. Electricity (0.13) and operation and maintenance (0.08) scores are modest, reflecting the system's balanced operational demands. Low $COD_t$ (17.92-134 mg/L) and $BOD_5$ (27-31 mg/L) levels explain favorable scores for CODt (0.08) and $BOD_5$ (0.06), indicating effective organic matter removal. Exceptional TSS performance (0.88-12.3 mg/L) corresponds to the very low TSS score (0.01), highlighting strong solids reduction. The low $NH_4$-N (0.07-4.82 mg/L) and TP (0.08-0.9 mg/L) concentrations are supported by excellent scores for $NH_4$-N (0.01) and TP (0.01), reflecting efficient nutrient removal. The moderate HRT range (0.5-1.8 days) corresponds to the reasonable HRT score (0.10), indicating a well-balanced hydraulic retention time that supports treatment efficacy. These insights collectively explain the high ranking of multi-soil-layering systems.

### SAND FILTER (TNS: 0.1907, RANK: 6)

Sand filter systems, with a TNS of 0.1907 and ranked sixth, display mixed performance. Low capital investment (0.06) and replacement (0.06) scores align with the relatively simple structure of sand filters. However, higher operation and maintenance (0.23) and electricity (0.13) scores reflect the need for frequent cleaning and energy input. The high $COD_t$ (15-886.91 mg/L) and BOD5 (3.6-438.75 mg/L) concentrations in some studies explain the poor scores for $COD_t$ (0.47) and $BOD_5$ (0.47), indicating inconsistent organic matter removal. TSS performance is weak, with values ranging from 11 to 830 mg/L, aligning with the high TSS score (0.60), suggesting limited solids filtration in certain setups. Moderate $NH_4$-N values (7.66-40 mg/L) justify a better $NH_4$-N score (0.10), reflecting its reasonable nutrient removal capacity. Lower TP concentrations (1.39-4.9 mg/L) result in a decent TP score (0.07). The HRT score (0.07) corresponds to the moderate retention times (0.35-1.27 days), indicating a balanced flow through the system. Overall, sand filters rank lower due to variable pollutant removal efficiency across different studies.

### ROTATING BIOLOGICAL CONTACTORS (RBCS) (TNS: 0.1229, RANK: 4)

Rotating Biological Contactors (RBCs), with a TNS of 0.1229 and ranked fourth, show a strong performance across most criteria. Low capital investment (0.09) and replacement (0.07) scores reflect the system's relatively simple design. The higher electricity (0.27) and operation and maintenance (0.23) scores indicate moderate energy needs and upkeep. $COD_t$ levels (45-105.3 mg/L) justify the low $COD_t$ score (0.08), indicating efficient organic matter removal. The $NH_4$-N score of 0.00 aligns with the extremely low $NH_4$-N values (0.02-0.08 mg/L) in the effluent, showcasing exceptional ammonia removal. The TP score (0.04) corresponds to the minimal phosphorus concentrations (0.15-0.21 mg/L), reflecting effective nutrient management. The HRT score of 0.00 aligns with the short retention times (0.36-0.63 days), enhancing process efficiency. Overall, RBCs balance cost and performance effectively, particularly excelling in nutrient removal.

### MOVING BED BIOFILM REACTORS (MBBRS) (TNS: 0.1013, RANK: 3)

Moving Bed Biofilm Reactors (MBBRs), with a TNS of 0.1013 and ranked third, show strong performance across various criteria. Low capital investment (0.06) and replacement (0.07) scores align with their moderate infrastructure requirements. Electricity (0.20) and operation and maintenance (0.23) scores indicate that MBBRs demand moderate energy and upkeep compared to simpler systems. The $COD_t$ score (0.05) reflects effective organic matter removal, supported by low $COD_t$ concentrations (23.64-75 mg/L) in the effluent. Similarly, the $BOD_5$ score (0.06) aligns with typical $BOD_5$ values (25-32 mg/L), showing strong biodegradation. Excellent TSS performance (15-22 mg/L) is reflected in a low TSS score (0.03). $NH_4$-N concentrations (2.11-10 mg/L) justify a strong $NH_4$-N score (0.02), demonstrating efficient ammonia removal. TP values (0.23-2.1 mg/L) explain the favorable TP score (0.03), indicating good phosphorus management. The HRT score (0.07) corresponds to short retention times (0.33-1.2 days), enhancing the system's operational efficiency. These factors make MBBRs highly effective, particularly in nutrient removal and organic matter reduction.

**DOWNFLOW HANGING SPONGE (DHS) FILTER (TNS: 0.0873, RANK: 1)**

DHS filters, ranked first with a TNS of 0.0873, exhibit outstanding performance across multiple criteria. Low capital investment (0.07) and replacement (0.07) scores reflect the system's affordability and durability. The moderate electricity (0.13) and operation and maintenance (0.08) scores align with its energy efficiency and low operational demands. The low $COD_t$ (32-69 mg/L) and $BOD_5$ (2-30 mg/L) values justify the excellent $COD_t$ (0.04) and $BOD_5$ (0.03) scores, indicating effective organic matter reduction. The TSS score (0.03) aligns with TSS concentrations (8-32 mg/L), showcasing strong solid removal capacity. While $NH_4$-N concentrations (28-100 mg/L) are higher, the corresponding $NH_4$-N score (0.26) reflects the challenge in ammonia reduction. The HRT score (0.02) aligns with short retention times (0.06-0.35 days), indicating rapid processing without compromising treatment quality. These factors explain DHS filters' top rank, combining cost-efficiency with strong pollutant removal capabilities.

## 4.2  INSIGHTS ON THE DHS FILTER, MSL, AND MBBR IN THE PHILIPPINES

Using a two-factor ANOVA without replication at a 0.05 level of significance, it was found that there is a significant difference across all the wastewater treatment technology types, with a p-value of 0.00486. However, when narrowing the analysis to only the top three technology types—namely, the Downflow Hanging Sponge (DHS) filter, Multi-Soil Layering (MSL), and Moving Bed Biofilm Reactors (MBBRs)—no significant difference was found between them, as evidenced by a p-value of 0.94735. This statistical analysis indicates that, despite their different operating principles and configurations, there is no strong evidence to suggest one is consistently superior to the others in the context of overall performance. Each technology offers distinct advantages, making them appropriate for specific operational needs.

DEWAT technologies are essential to address the wastewater management challenges in the Philippines, particularly those consequent to its geographical layout and limited access to centralized infrastructure. In this context, DHS filters, MSL systems, and MBBRs stand out as particularly relevant for local adaptation. The DHS filter, known for its energy efficiency, operates using gravity flow with minimal energy input (Tyagi et al., 2021). This feature is especially advantageous in remote or off-grid areas in the Philippines where energy resources may be scarce. Conversely, MBBRs require substantial energy for aeration, making them potentially less suitable for such environments. However, the DHS filter may necessitate periodic maintenance to manage sponge media clogging, though this is less of an issue in MBBRs, which use durable plastic carriers, or MSL systems, which rely on natural layer compositions. Extensive experimental evidence from laboratory and pilot-scale studies shows that DHS filters can offer long periods of operation with minimal maintenance (Maharjan et al., 2020), making them practical for deployment in rural or hard-to-reach areas across the Philippines. Some studies even suggest DHS filters can maintain a sludge retention time of up to 100 days without significant clogging (Maharjan et al., 2020; Nasr et al., 2022; Tyagi et al., 2021), which is significantly longer than what is typically reported for MSL systems and MBBRs (Ho & Wang, 2015; Latrach et al., 2014; Saidulu et al., 2021; Zhang et al., 2015). This extended retention time is partially due to the sponge material's ability to support diverse autotrophic and heterotrophic microorganisms (Hatamoto et al., 2018; Kubota et al., 2014) that efficiently degrade particulate organics (Onodera et al., 2013). The success of sponge material in cultivating diverse microbial communities, including previously uncultured microorganisms (Imachi et al., 2022), makes the DHS filter highly effective in treating various contaminants, further supporting its application in decentralized systems in the Philippines.

MSL systems, on the other hand, offer a sustainable and environmentally friendly approach by using a configuration of soil and organic materials capable of removing a broad spectrum of contaminants, including nutrients and heavy metals (Latrach et al., 2014). This makes them valuable in regions where contamination from agricultural runoff or mining is a concern, both of which are critical issues in certain areas of the Philippines. However, the larger spatial footprint required for MSL systems may pose a limitation in densely populated urban centers or in areas with limited land availability. Furthermore, MSL systems present potential clogging risks, which could require more frequent maintenance compared to the compact designs of both DHS filters and MBBRs.

MBBRs, in contrast, are known for their compact size and operational flexibility, offering high treatment efficiency under varying conditions (Atasoy et al., 2007; Jabornig & Favero, 2013; Saidulu et al., 2021). This makes them well-suited for areas experiencing fluctuating population densities, such as seasonal tourist destinations in the Philippines. Additionally, MBBRs have the advantage of low sludge production and ease of scalability (Saidulu et al., 2021), making them attractive for decentralized applications where future expansion may be necessary. While MBBRs require more energy, their scalability and ease of operation under different loading conditions can make them viable in specific contexts within the Philippines, particularly in industrial or semi-urban areas where access to energy is more reliable.

The modular design of the DHS filter offers a distinct advantage in decentralized wastewater treatment systems (Tyagi et al., 2021). This modularity allows for easy retrofitting and scaling of systems, making it highly adaptable for DEWAT applications across the diverse landscapes of the Philippines, from rural villages to coastal towns. Its ability to be retrofitted onto existing systems offers a cost-effective solution for areas that cannot afford entirely new infrastructure, addressing the financial constraints often faced by developing countries like the Philippines (Nasr et al., 2022). In contrast, the MSL system's larger footprint and potential for clogging may limit its use in such settings, even though it offers a broader contaminant removal spectrum.

While the statistical analysis shows no significant performance difference between the DHS filter, MSL systems, and MBBRs, each technology has strengths that align with different operational needs. For a developing country like the Philippines, the decentralized nature of these systems offers a path forward in improving wastewater management, with DHS filters being particularly advantageous in remote or resource-scarce environments, MSL systems offering broader contaminant removal in agricultural and mining regions, and MBBRs providing flexibility and scalability in semi-urban or industrial zones. These technologies, when implemented appropriately, can contribute to the country's overall efforts toward sustainable wastewater management and environmental protection.

## 5 CONCLUSION

In conclusion, this study presents a comprehensive evaluation of decentralized wastewater treatment (DEWAT) technologies using a multi-criteria analysis (MCA) approach, specifically employing the Analytic Hierarchy Process (AHP) and the Delphi method. The assessment highlights that the Downflow Hanging Sponge (DHS) filter, Multi-Soil Layering (MSL) system, and Moving Bed Biofilm Reactor (MBBR) are among the top-ranked technologies suitable for the Philippine context. These three technologies showed no statistically significant differences in overall performance, but each possesses unique characteristics that align with specific operational requirements, demonstrating their potential adaptability to various local contexts.

The DHS filter emerges as the leading choice due to its energy efficiency, modularity, and cost-effectiveness, which make it highly suitable for deployment in rural and remote areas. Its extended sludge retention time and minimal maintenance requirements position it as an optimal solution for off-grid locations, particularly where maintenance resources are scarce. The MSL system, on the other hand, offers a sustainable solution with strong nutrient removal capabilities, making it particularly valuable in agricultural. Data from literature suggest that it can be great for heacy metal in mining areas, which are prevalent in the Philippines. However, its larger footprint and potential clogging risks may limit its use in densely populated regions. Meanwhile, the MBBR offers compactness, scalability, and high efficiency under varying conditions, making it suitable for semi-urban or industrial zones, especially where access to energy is reliable.

The findings of this study underscore the importance of context-specific considerations in selecting appropriate DEWAT technologies. While the DHS filter stands out for rural areas, MSL systems are ideal for regions dealing with specific types of contamination, and MBBRs are well-suited for dynamic environments with fluctuating wastewater loads. The modular and scalable nature of these technologies makes them adaptable for decentralized applications across the diverse landscapes of the Philippines, providing viable alternatives to centralized wastewater management systems that are often impractical in the country.

Ultimately, the adoption of DEWAT technologies like the DHS filter, MSL system, and MBBR offers a promising pathway to addressing the wastewater management challenges in the Philippines. These technologies align well with the country's goals for sustainable environmental management, providing efficient, adaptable, and cost-effective solutions. Future research can explore the long-term performance of these technologies for DEWAT applications under varying environmental conditions, particularly in tropical settings like the Philippines. Of great research interest would be applications in remote/off-grid islands and similar settings that have limited energy access. Additionally, studies on optimizing maintenance protocols and other operational aspects of these technologies for DEWAT applications is suggested. Endeavoring to minimize the spatial footprint of systems like MSL could enhance their applicability in densely populated areas. Mathematical modeling research on these technologies, which is observed to be minimal, should promote their optimization as these models offer a way to simulate them thereby reducing the cost of optimization. Further analysis of hybrid systems combining multiple technologies may also lead to more efficient wastewater treatment solutions tailored to diverse operational contexts.

## 7 ACKNOWLEDGEMENTS

The present work is funded by the Engineering Research and Development for Technology (ERDT) scholarship of the Department of Science and Technology (DOST), Republic of the Philippines. The useful conversations with Engr. Luis K. Cabatingan from the Department of Chemical Engineering, which inspired EFSJ to pursue with his dissertation topic is also acknowledged.

## 8 DATA AVAILABILITY STATEMENT

The step-by-step procedures of the MCA, its raw data, as well as the calculation is available in a supplementary file accessible through the following link: https://doi.org/10.17632/wgycpw7byv.1

The details of the ANOVA results on the MCA dataset, i.e., ANOVA table and calculations, is available in a supplementary file accessible through the following link: https://doi.org/10.17632/wgycpw7byv.1

## 9 CONFLICT OF INTEREST

The author declares no conflict of interest.